\documentstyle[osa,psfig]{revtex}

\newcommand{\be}{\begin{equation}}
\newcommand{\ee}{\end{equation}}
\newcommand{\nn}{\mbox{} \nonumber \\ \mbox{} }
\newcommand{\ba}{\begin{eqnarray}}
\newcommand{\ea}{\end{eqnarray}}

\newcommand{\Alfven}{ Alfv\'{e}n }

\begin{document}


\title{Dynamics of relativistic magnetized blast waves}
\author{Maxim Lyutikov }
\address{McGill University, 3600 rue University
Montreal, QC, H3A 2T8   \\ and \\
Massachusetts Institute of Technology,
77 Massachusetts Avenue, Cambridge, MA 02139
\footnote{CITA National Fellow}
}
\maketitle

\date{Received   / Accepted  }

\begin{abstract}
The dynamics of  a
relativistic blast wave  propagating through magnetized  
medium
is considered  
 taking into account   possible inhomogeneities 
 of density and magnetic field
 and additional energy supply.
Under the simplifying assumption of a spherically symmetric explosion
in a medium with toroidal magnetic
field  self-similar solutions for the
internal dynamics of the flow are derived.   
In the weakly magnetized case, when  the 
bulk of the flow may be described by the unmagnetized solutions, there
is a strongly magnetized sheath near the  contact
discontinuity (when it exists).  
Self-similar solutions inside the sheath are investigated.
In the  opposite limit of 
strongly magnetized upstream plasma new analytical
 self-similar solutions are found.
Possible application to the physics of Gamma-Ray Bursts 
 is discussed.
\end{abstract}

\section{Introduction}

Recent observational advances in the area of Gamma-Ray Bursts (GRBs) 
(see, e.g., Ref \onlinecite{M2001} for a review) have
stimulated  research on the physics of  strongly relativistic explosions.
Most analytical results for the dynamics of strong explosions
belong to the class of self-similar solutions. This approach
 allows  great simplification
in reducing a system of partial differential equations to the ordinary ones
and often represents an asymptotic behavior of the flow.
Following the  seminal works of Sedov \cite{Sedov}
(for non-relativistic strong explosions) and  of  Blandford \& McKee  \cite{BMc}
(B\&M hereafter)
a series of generalizing works have been done. 
Non-relativistic shock waves propagating in  a magnetized medium have been
considered \cite{LeeChen,Rosenau76,Summers75} as well as 
generalizations of the B\&M solutions 
 for a wide class of 
density profiles
\cite{WaxmanShvarts,BestSari}.
 Yet, until now no account of the
dynamics of the 
magnetic field in relativistic blast waves has been made.   
A lack of  the relativistic treatment of the
blast wave in magnetized medium is noticeable, especially in   view
of heuristic interpretation of GRB emission as due to  
synchrotron emission in  strongly magnetized relativistic blast waves
(e.g., Ref \onlinecite{Piran00}).
This is done in the present work. In a follow-up work  \cite{Lyutikov01}
 we will explore 
the internal structure of the relativisticly expanding magnetic cavity.

\section{Formulation of the problem}

We seek self-similar solutions to the
relativistic dynamics of a  spherical  expansion  of gas into a magnetized 
medium due to strong explosion. 
We assume that
(i) time dependent  energy source is located in the center;
(ii)   the  magnetic field
is  toroidal and  spherically symmetric;
(ii) there is a cold  spherically symmetric external medium with density $n_1$
(which may depend on radius).

 The assumption of spherical symmetry with toroidal 
magnetic field is a controversial, but a
commonly employed simplification. A number of works 
applied spherical approximation
to MHD  winds from  pulsars
 \cite{Michel69,KennelCoroniti84,EmmeringChevalier87} and to  
explosions in  the Solar wind
 \cite{Rosenau76,Summers75}.
Other astrophysical setting  where such situation may arise is 
an explosion in 
 a preexisting cavity blown out by 
spherical magnetized stellar wind.
Formally, our approach also applies  to the equatorial region of 
an explosion in a constant magnetic field, where the shock velocity
is  perpendicular to the magnetic field.

The formal treatment of the problem, in which we rely mainly on the
 approach of B\&M, starts with the 
set of relativistic magnetohydrodynamic equations which 
can be written in terms of conservation laws \cite{LandauII}
\ba && 
T^{ij}_{,i} =0,
\label{x1}
 \\ &&
F^{\ast\, ij}_{,i} =0,
\label{x2}
\\ &&
(\rho u^i)_{,i}=0
\label{x3}
\ea
where
\be
T^{ij} = (w +b^2) u^i u^j +(p+{1\over 2} b^2) g^{ij} -b^i\,b^j
\ee
is the energy-momentum tensor, $w= 4  p $ is 
 plasma proper enthalpy (upfront in the cold plasma
$ w_1 =n_1 m_i c^2$, 
where $n_1$ is external density, $m_i$ is the ion mass, while
behind the shock  $w = 4 p$ appropriate for relativistic plasma),
$b^2 = b_i b^j$ is the plasma proper magnetic energy density times $4 \pi$, 
$p$ is pressure, $u^i=(\gamma, \gamma {\bf \beta})$ 
are the plasma four-velocity, 
Lorentz-factor and three-velocity, $g^{ij}$ is the metric tensor, 
$b_i = {1\over 2} \eta_{ijkl} u^j F^{kl}$ are the four-vector of magnetic
field, Levy-Chevita tensor and electro-magnetic field tensor.
It has  also been  implicitly assumed in the derivation of 
these equations that one of the electromagnetic invariants is not
equal to 0 and  the electro-magnetic stress energy tensor can be
diagonalized  (equivalently,  this implies that there
is a  reference frame where the magnetic (or electric) field is not equal to
0). 

Writing  out Eqns (\ref{x1}-\ref{x3}) in coordinate form and 
assuming a spherically symmetric outflow with toroidal magnetic field,
the conservation of energy and momentum (\ref{x1}), induction equation 
(\ref{x2}) and mass conservation  give
\ba 
&&
\partial_t \left[ ( w+ b^2) \gamma^2 -(p+b^2/2)\right]+
{1\over r^2} \partial_r \left[ r^2 ( w+b^2) \beta \gamma^2 \right] =0
\label{x4} \\ &&
\partial_t \left[ ( w+ b^2) \gamma^2 \beta \right]+
{1\over r^2} \partial_r 
\left[r^2\left((w+b^2) \beta^2 \gamma^2 + (p+b^2/2)\right) \right] - 
{2 p \over r}=0
\label{x5} \\ &&
\partial_t \left[  b \gamma \right]+
{1\over r} \partial_r  \left[r b \beta \gamma \right] =0
\label{x51} \\ &&
\partial_t \left[ \rho \gamma \right]+
{1\over r^2 } \partial_r  \left[r^2 \rho  \beta \gamma \right] =0
\label{x6}
\ea

Equation (\ref{x4}-\ref{x6}) should be complemented by the boundary conditions
on the shock front.
The jump conditions on the relativistic  transverse magnetized shocks
may be written as \cite{KennelCoroniti84}
\ba
&&
\gamma^2 = \frac{\gamma ' - u' }{\gamma '+ u'} \Gamma^2 =
\Gamma^2 \times { \left\{ \begin{array}{cc}
{1  \over 2} & \mbox{  if $ \sigma \ll 1$} \\
{ 1 \over 4  \sigma} & \mbox{  if $ \sigma \gg  1$}
\end{array} \right.}
\nn &&
p_2 = \frac{m\,{n_1}\, c^2 } {4\,{u}'\,
    {{\gamma }}'} {
\,\left( 1 +
      \sigma \,\left( 1 - \frac{{{\gamma }_2}'}{{u_2}'} \right)  \right) }
    \, {\Gamma }^2 =
n_1 m c^2 {\Gamma }^2  \times \left\{ \begin{array}{cc}
{ 2 \over 3} & \mbox{  if $ \sigma \ll 1$} \\
{ 1 \over 8  \sigma} & \mbox{  if $ \sigma \gg  1$}
\end{array} \right.
\nn &&
n_2= { n_1 \Gamma \over u'}= n_1 \Gamma  \times \left\{ \begin{array}{cc}
2 \sqrt{2} & \mbox{  if $ \sigma \ll 1$} \\
{1 \over \sqrt{\sigma}} & \mbox{  if $ \sigma \gg  1$}
\end{array} \right.
\nn &&
b_2= {\Gamma \over u'} b_1 = { \Gamma \over u'} \sqrt{ \sigma n_1 m c^2} =
\sqrt{ n_1 m c^2} \Gamma \times \left\{ \begin{array}{cc}
2\,\sqrt{2 \,\sigma }  & \mbox{  if $ \sigma \ll 1$} \\
1 & \mbox{  if $ \sigma \gg  1$}
\end{array} \right.
\label{bo}
\ea
where
\be
{u'}^2= \frac{1 + 10\,\sigma  + 8\,{\sigma }^2}{16\,\left( 1 + \sigma  \right) }
 +
  \frac{{\sqrt{1 + 20\,\sigma \,\left( 1 + \sigma  \right)  +
        64\,{\sigma }^2\,{\left( 1 + \sigma  \right) }^2}}}
{16\,\left( 1 + \sigma  \right) }
, \hskip .3 truein
{\gamma ' }^2 = {u'}^2 +1
\ee
are the four speed and Lorentz factor of the shocked fluid 
in the frame of the shock
and
\be
\sigma = \frac{{{b_1}}^2}{m\, c^2\,{n_1}}
\ee
is the magnetization parameter in front of the shock.
The subscripts denote the quantities in the unshocked ($1$) and 
shocked ($2$) media.

Relations (\ref{bo}) allow for the following 
  parameterization
\ba &&
p_2 = \Gamma^2 n_1 m c^2 f(\chi)
\nn &&
\gamma^2 =  \Gamma ^2 g(\chi)
\nn &&
n_2 =  \Gamma n_1 n(\chi)
\nn &&
b_2 = \Gamma  b_1 h(\chi)
\ea
with the boundary conditions
$ f(1) =f_0$, $ g(1) =g_0$,$ h(1) =h_0$, $ n(1) =n_0$ (see below).
\footnote{We choose to  work consistently with proper quantities, 
i.e. measured in the plasma
rest frame. One should be careful in comparing our equation with
B\&M and Ref. \onlinecite{KennelCoroniti84}}

Following   B\&M
we choose the self-similar variable
\be
\chi = 1+2(m+1) \xi =[1+2(m+1) \Gamma^2](1-r/t)
\ee
where $\xi = (1-r/R)  \Gamma^2$, $\Gamma$ is the Lorentz factor of the
shock, 
$R= t \left( 1-1/(2(m+1) \Gamma^2)\right)$ is the radius
 of the contact discontinuity and 
we assume  the Lorentz factor scales with radius as
$\Gamma^2 \propto t^{-m}$.
We limit ourselves to the strongly relativistic case 
expanding all relations  to  first order in $1/\Gamma^2$.

Treating $(\chi,y)$, where $y= \Gamma^2$, as new independent variables we find
\ba&&
\partial_ t = - m y \partial_{y} +  ((m + 1)( 2 y - \chi) +1) \partial_{\chi}
\nn
&&
\partial_r = -  (  1 +  2(m + 1) y)  \partial_{\chi}
\nn &&
\beta = 1-{1 \over 2 y g}
\nn &&
r=  t \left( 1-{\chi \over 1+2(m+1) y}\right)
\ea

The equations for the self-similar 
variables
$f$, $g$, $n$ and $h$ read
\ba
&&
{{\cal{A}} \over g} { \partial \ln f \over  \partial \chi}= 
- 2\,\left( 4 (1 - m) + \left(m -4 \right) \,\chi\, g \right)  \, f+ 
   \frac{ 2(1 - m ) + 3\,\left( m -2 \right) \,\chi  \, g   }{
    \left(  \chi\, g -1 \right) } \, h^2
\nn &&
{{\cal{A}} \over g} { \partial \ln g \over  \partial \chi}= 
 - 
\left( 4 - 7\,m + 2\,\left( 2 + m \right) \,\chi \, g \right)  \, f  +
{3\,\left(m -1  \right) } \,h^2
\nn &&
{{\cal{A}}\over g} { \partial \ln h  \over  \partial \chi}= 
 -\frac{\left( 2\,\left(m -4  \right) \,{\chi }^2 \, g^2 + 
   \left( 8 - 11\,m \right) \,\left(  \chi\, g  -1 \right)  \right) }{2\,
   \left( \chi\, g -1 \right) } \, f + \frac{3\,\left(m -1\right) }{2 } \,h^2
\nn &&
{{\cal{A}}\over g} { \partial \ln n \over  \partial \chi}=
{1\over 2(  \chi\, g  -1)} \left(
- 
  {f\,\left( -12 + 11\,m + g\,\left( 24 - 11\,m \right) \,\chi  + 
       2\,g^2\,\left(m -6  \right) \,{\chi }^2 \right) } +
\right.
\nn &&
\left. 
{3\,h^2\,\left( 1 - m + g\,\left( m-3  \right) \,\chi  \right) } \right)
\nn &&
{\cal{A}}  = \left( 1 + m \right) \,
\left(  2\,f\,\left( 1 - 4\, \chi\, g  + g^2\,{\chi }^2 \right) -3\, \chi \,g\, h^2  \right)
\label{15}
\ea
System (\ref{15}) is a relativistic generalization of Eqns. (3.10-3.13) of 
Ref. \onlinecite{Summers75}
In the limit $ h \rightarrow 0$ 
these relations  reproduce the unmagnetized case of B\&M.
\footnote{
The system (\ref{15}) may in fact  be transformed into a system
of one ODE and three equations resolvable in quadratures if we 
change to a new coordinate $x=  \chi\, g$ and introduce a magnetization
parameter $\beta = f /h^2$.}

Sometimes it is  more convenient to use 
 use the  functions $\tilde{g}, \tilde{f}, \tilde{n} 
, \tilde{h} $  and $\tilde{x}= \tilde{g} \chi$
with  boundary conditions on the shock front
$ \tilde{g}=  \tilde{f}=  \tilde{n} = \tilde{h}  = \tilde{x}= 1$:
\ba &&
g= g_0 \tilde{g}, \hskip .2 truein 
f= f_0 \tilde{f}, \hskip .2 truein 
n= n_0 \tilde{n}, \hskip .2 truein
h= n_0 \tilde{h}, \hskip .2 truein
x\equiv g \chi = g_0 \tilde{x}
\nn &&
 g_0 = \frac{u' + \gamma '}{\gamma '- u'}
\approx \left\{
\begin{array}{ll}
{1 \over 2} & \mbox{ $\sigma \ll 1$} \\
{ 1 \over 4 \sigma}  & \mbox{ $\sigma \gg  1$}  
\end {array} \right.
\nn &&
 f_0 = \frac{1}{4 \,{u_2}\, {{\gamma }_2}' }
\left( 1 +
      \sigma \,\left( 1 - \frac{{{\gamma }_2}'}{{u_2}'}   \right)  \right)=
\left\{
\begin{array}{ll}
 {2\over 3} & \mbox{ $\sigma \ll 1$} \\
{1\over 8 \sigma} 
 & \mbox{ $\sigma \gg  1$}
\end {array} \right.
\nn &&
n_0 = { 1\over u'} \approx 
\left\{
\begin{array}{ll}
2  \sqrt{2}  & \mbox{ $\sigma \ll 1$} \\
{ 1 \over \sqrt{\sigma} } 
& \mbox{ $\sigma \gg  1$}
\end {array} \right.
\nn &&
h_0 = {\sqrt{ \sigma } \over u'} \approx
\left\{
\begin{array}{ll}
 2  \sqrt{ 2 \sigma } & \mbox{ $\sigma \ll 1$} \\
1 & \mbox{ $\sigma \gg 1$}
\end {array} \right.
\label{L}
\ea

\begin{figure}[h]
\psfig{file=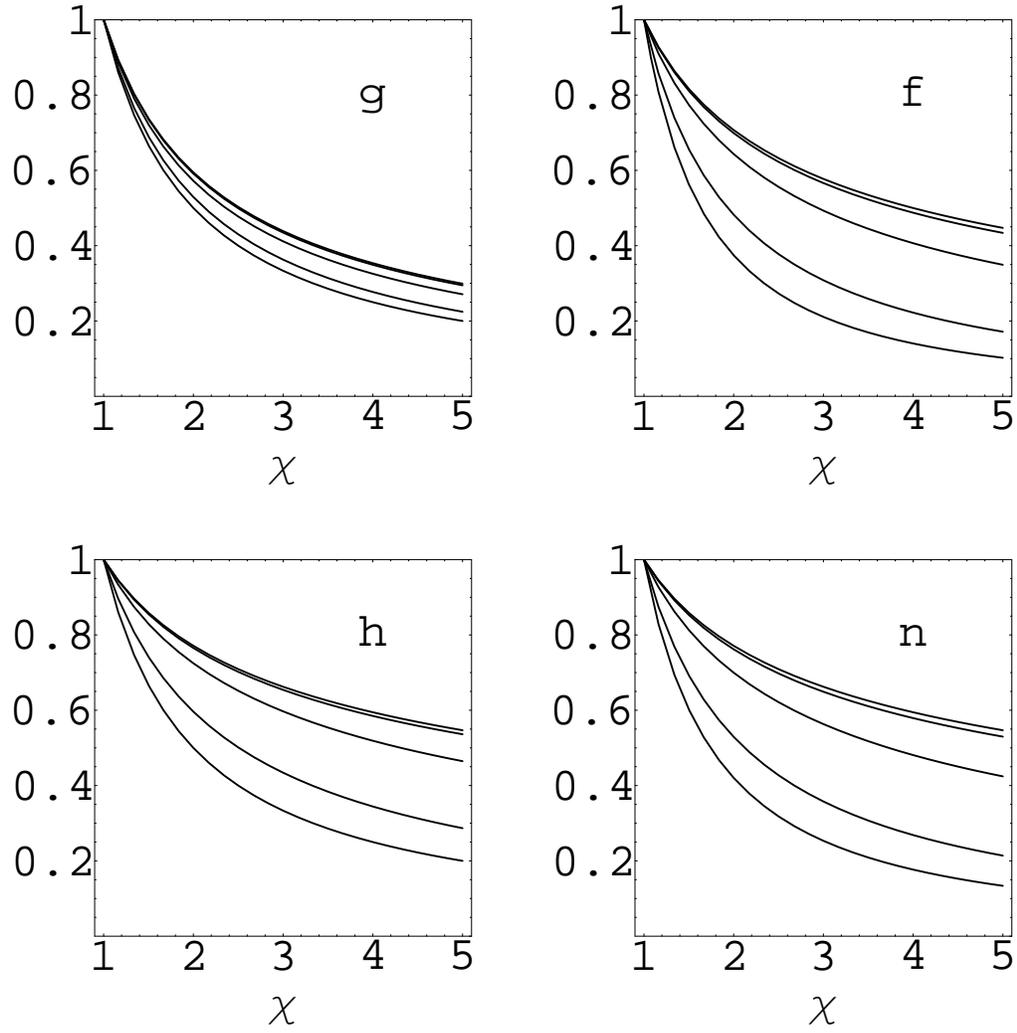,width=15cm}
\caption{Lorentz factors, pressure, magnetic field and density for
$m=3$. The curves from top to bottom are (i) asymptotic limit
$\sigma \gg 1 $, (ii) $\sigma=10$, (iii) $\sigma=1$, (iv) $\sigma=0.1$
and $\sigma=0$. In the case $\sigma =0$ the magnetic field is normalized to
some arbitrarily small initial value.
}
\label{m3sigma}
\end{figure}

\begin{figure}[h]
\psfig{file=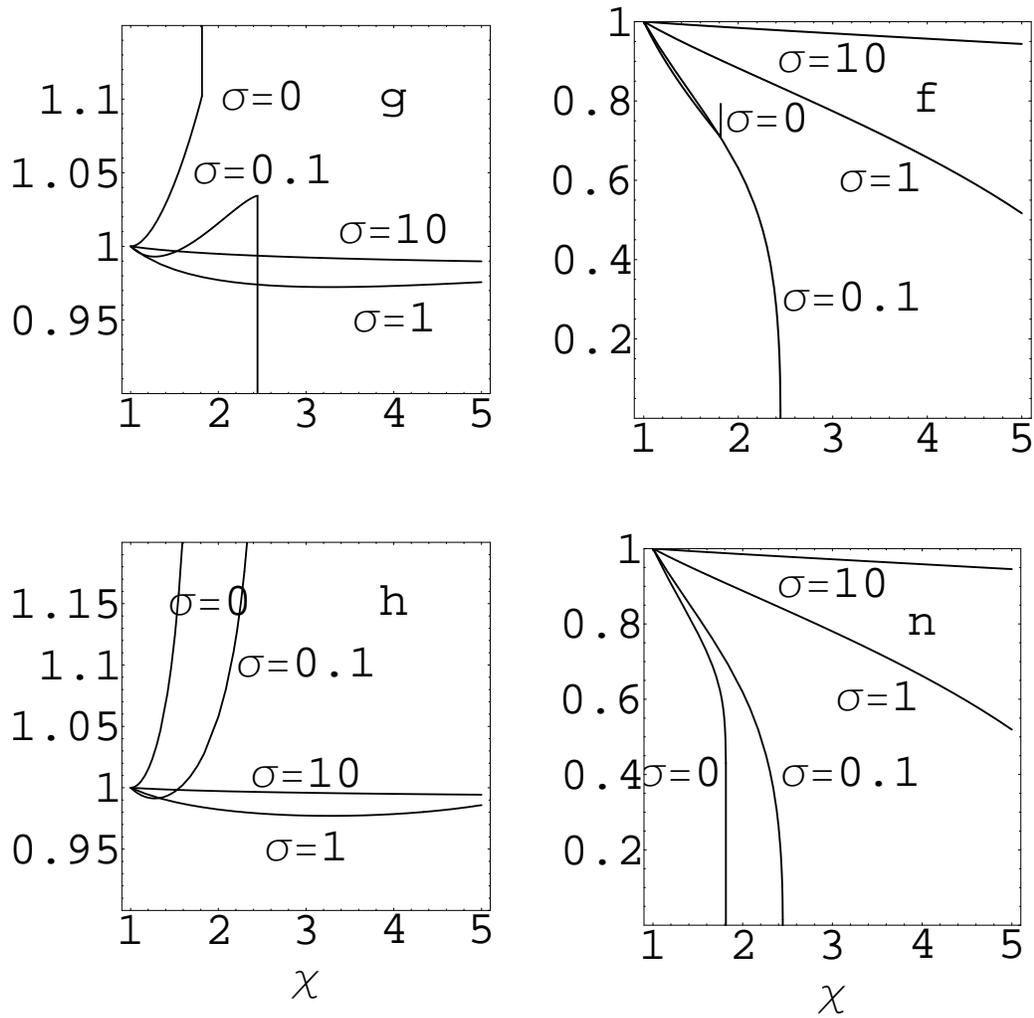,width=15cm}
\caption{Lorentz factors, pressure, magnetic field and density for
$m=1$. The curves are labeled by the values of $\sigma$.
For $\sigma \gg 1$ the asymptotic solutions are
$g=f=h=n=1$. Note that magnetic field piles up on the contact discontinuity.
}
\label{m1sigma}
\end{figure}

We can solve the above equations by direct numerical integration.
(see Fig. \ref{m3sigma} and \ref{m1sigma}).
Simple analytical relations, though, can be obtained in the
weakly, $\sigma \ll 1$, and strongly magnetized, $\sigma \gg 1$, limits.

\section{Small $\sigma$ limit}

Relations for $\tilde{g},\tilde{f}$ and $\tilde{n}$ 
in this case are the same as for the
unmagnetized case (B\&M). For the evolution of the overall passive
 magnetic field we find
\be
{1\over \tilde{g}} {\partial \ln \tilde{h} \over \partial \chi}=
-\frac{\left( -16 + 22\,m + \left( 8 - 11\,m \right) \,\chi \tilde{g} + 
      \left(m-4 \right) \,{\chi }^2 \, \tilde{g}^2 \right) }{2\,\left( 1 + m \right) \,
    \left( \chi \, \tilde{g} -2  \right) \,\left( 4 - 8\,\chi \, \tilde{g}  
+ {\chi }^2 \, \tilde{g}^2 \right) }
\label{ql}
\ee
which can be integrated using the known solution for $\tilde{g}$.
In particular, for $m=3$, this can be resolved for $\tilde{h}(\chi)$:
\be
\tilde{h}(\chi) = { 1 \over \chi}
\ee
Recall that in this case $\tilde{g}= 1/\chi$, $\tilde{f} = 1/\chi^{17/12}$ 
and $ \tilde{n} =  1/\chi^{5/4}$. 
The magnetization parameter, $\tilde{\beta}= \tilde{f} / \tilde{h}^2 = 
1/\chi^{5/12}$ in this case is a decreasing function of $\chi$,
 so if it is small
on the shock it will always remain small.

\subsection{Magnetosheath}

When the energy contained inside the shock increases with time a
power source is necessary. This can be an expanding piston or another fluid
with a larger Lorentz factor. In both cases a contact discontinuity (CD) forms
separating the external and internal fluids.
As discussed by B\&M the point $\chi \, g \equiv x = 1$ is the 
location of the contact discontinuity.
As Eq. (\ref{ql}) (which is the small $\sigma$ limit of exact
relations) suggests, the magnetic field increases infinitely near the
CD. In reality, of course,  it will grow until approximate equipartition
is reached
forming a thin boundary layer.  This is consistent with the
non-relativistic consideration  \cite{Kulsrud,Rosenau76}
where it was   noted that   however small the
magnetic field is in the unshocked medium near the CD the magnetic
field grows in magnitude and starts  to dominate the dynamics
of the  thin boundary layer.  In this section we study the dynamics
of the magnetosheath.

To estimate the thickness of the magnetosheath
we use the system  (\ref{5}) using  independent variable $x$
($k=l=0$ in this section).
In the small $\sigma$ limit the local magnetization parameter 
$\beta = {h^2 / f}$ is
\be
\beta   \sim   (x-1)^{2(m-4)/(12-m)}
\ee
Since near the CD $ g\sim {\rm constant}$ the width of the
magnetic sheath in coordinates $x$ is 
\be
 \Delta x \propto \sigma ^{ (12 -m) / 2 (4- m)}  
\ee
The constant of proportionality here is a complicated function of $m$.
The width of the magnetic sheath in coordinates $\chi$ is 
\be
\Delta \chi = { \Delta x  \over g'_{CD}/g_{CD} + g_{CD}}
\ee
where 
$g'_{CD}$ and $g_{CD}$ are the values taken at the CD.

Next we study the structure of the boundary layer.
Using $x$ as independent variable and retaining leading terms
near $x=1$ we find
\ba
&&
  { \partial \ln f \over  \partial x }= -
\frac{h^2\,\left(m-4 \right) }
   { \left( 6\,h^2 - f\,\left( m-12 \right)  \right)  \,\left( x-1 \right) }
\nn &&
 { \partial \ln g\over  \partial x}= 
  \frac{3\,h^2\,\left( m-1 \right)  + f\,\left( -8 + 5\,m \right) }
   {-6\,h^2 + f\,\left( m-12 \right) }
\nn &&
 { \partial \ln h \over  \partial x}= 
\frac{f\,\left(m-4 \right) }
   {\left( 6\,h^2 - f\,\left( m-12 \right)  \right) \,\left( x-1 \right) }
\nn &&
 { \partial \ln n \over  \partial x} =
 \frac{\left( 2\,f\,m + 3\,h^2\,\left( -1 + 3\,x \right)  \right) }
   {2\,\left( 6\,h^2 - f\,\left( m-12 \right)  \right) \,\left( x-1 \right) }
\label{1q}
\ea
which  immediately gives
\be 
f+ h^2/2 =  C_1
\label{EE}
\ee
where $C_1$ determines the total energy flux through the magnetosheath.
For simplicity  below we put $C_1$ equal to 1.

Using  (\ref{EE}) we can integrate the equation for the magnetic field
\be
h^{m-12}\,{\left( h^2-2 \right) }^6\,{\left( x-1 \right) }^{m-4} = C_2
\label{dd}
\ee
where $C_2$ is a constant of integration.
Relation (\ref{dd}) implicitly determines  the self-similar
structure of the magnetic field
inside the boundary layer.

As we approach the CD, $x \rightarrow 1$,
\be
h \sim \sqrt{2 } - (x-1)^{(4-m)/6}
\ee
This
 shows that for $m<4$, as $x \rightarrow 1$ the magnetic field
goes to a constant on the CD:
 $h \rightarrow \sqrt{ 2 }$. Similarly we  find that
$f \sim  (x-1)^{(4-m)/6} \rightarrow 0$ 
and $ n  \sim  \sqrt{x-1}  \rightarrow 0$.
Thus,  pressure and density vanish at the CD while  the
magnetic field is finite.
This result confirms that  no matter how small the
external  magnetic field is, there will be  a region near the CD  where
magnetic field pressure is dominant.
This is qualitatively different from the
hydromagnetic case where (for $m>0$)  the density was vanishing on the CD,
 resulting
 in  very high temperatures (c.f. Ref \onlinecite{Rosenau76}). 

The assumption of ideal MHD should break down
 somewhere inside the boundary layer
where
diffusive and dissipative processes are likely to play a very important role. 

The solution inside the boundary  layer should be matched to the
solution in the bulk flow. The exact relations connecting
$  C_1$ and $  C_2$ to $\sigma$ and $m$ 
 are prohibitively complicated for reproduction here.

\section{Large  $\sigma$ limit}

Simple 
analytical results may be obtained in the case of a strong magnetization,
$ 1 \ll \sigma \ll \Gamma^2 /4 $. The upper limit on $\sigma$ comes from
the fact that for 
 $\sigma \geq \Gamma^2 /4 $ the shock is no longer strong: it's four-speed
become comparable to the upstream \Alfven four-velocity.

Changing  to tilde-functions and expanding Eq. (\ref{15})
 for $\sigma \gg 1$ we find
\ba
&&
{1 \over   \tilde{g}} 
{ \partial \ln  \tilde{h}  \over \partial \chi } =
\frac{3\,\tilde{h}^2\,\left(  m -1\right) }
  {2\,\left( 1 + m \right) \,\left( \tilde{f} - 3\,\tilde{h}^2\,\chi \, \tilde{g} \right) }
\nn  && 
{1 \over   \tilde{g}}
{ \partial \ln  \tilde{g}  \over \partial \chi } = 
\frac{3\,\tilde{h}^2\,\left(  m -1\right) }
  {\left( 1 + m \right) \,\left( \tilde{f} - 3\,\tilde{h}^2\,\chi \, \tilde{g
} \right) }
\nn  && 
{1 \over   \tilde{g}} 
{ \partial \ln  \tilde{n}  \over \partial \chi } =
\frac{3\,\tilde{h}^2\,\left(  m -1\right) }
  {2\,\left( 1 + m \right) \,\left( \tilde{f} - 3\,\tilde{h}^2\,\chi \, \tilde{g
} \right) }
\nn  && 
{1 \over  \tilde{g}} 
{ \partial \ln  \tilde{f}  \over \partial \chi } =
\frac{2\,\tilde{h}^2\,\left(  m -1\right) }
  {\left( 1 + m \right) \,\left( \tilde{f} - 3\,\tilde{h}^2\,\chi \, \tilde{g
} \right) }
\label{19}
\ea

Which immediately gives
\be
\tilde{g} = \tilde{h}^2, \hskip .3 truein
\tilde{f} = \tilde{h}^{4/3}, \hskip .3 truein
\tilde{n} = \tilde{h}
\label{other}
\ee
In the strongly magnetized limit the pressure $\tilde{f}$ is proportional to
density $\tilde{n}$ to the  $4/3$ power. 

\begin{figure}[h]
\psfig{file=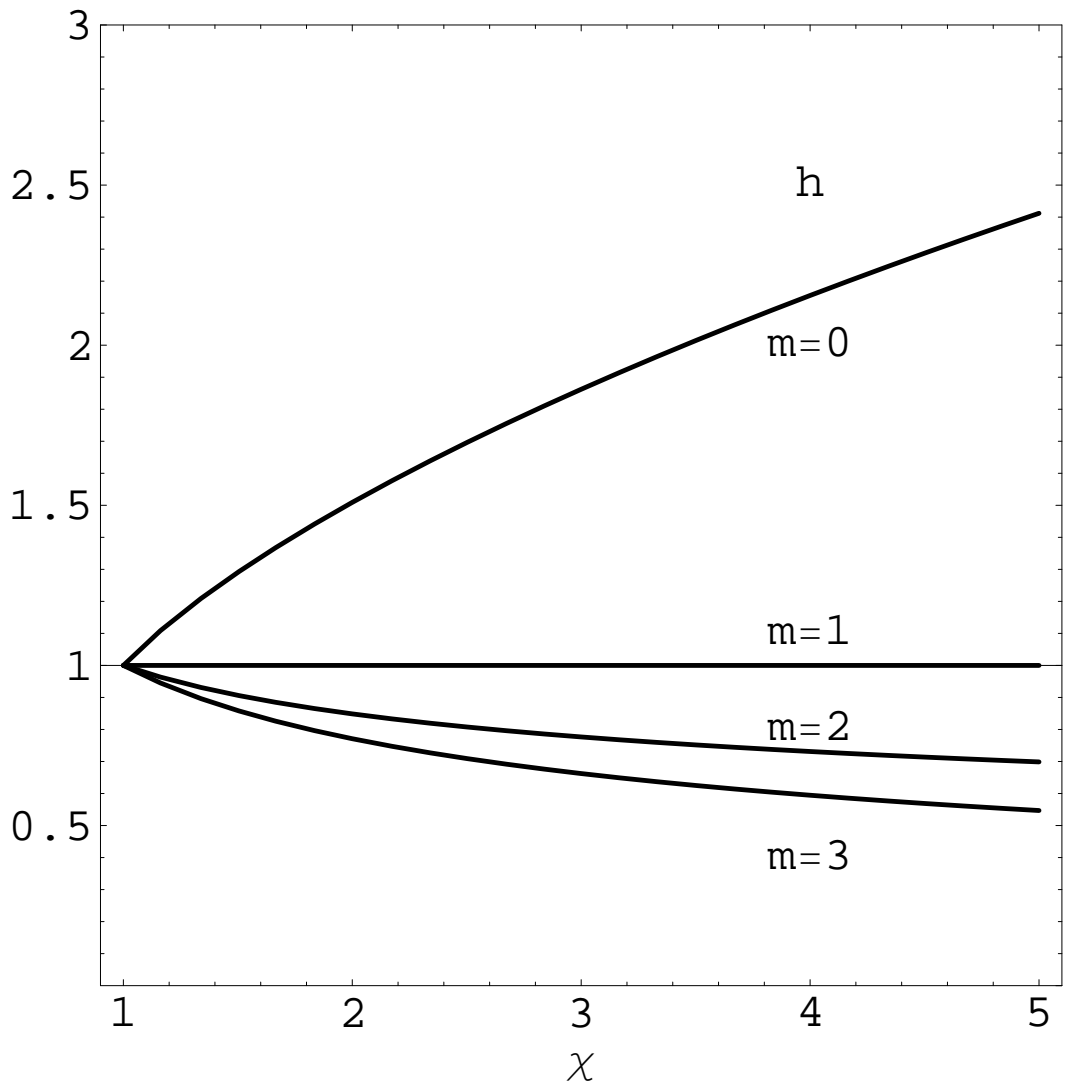,width=15cm}
\caption{ Magnetic field $\tilde{h}$
in the strongly magnetized limit $\sigma \gg 1$
for different values of $m$ ($k=l=0$).
 Other functions  are powers of $\tilde{h}$
(eq. \ref{other}).
}
\label{hofchi}
\end{figure}

We can then resolve the equations for $\tilde{h}$ as an implicit function
of $\chi$:
\be
\chi = \frac{2\,\left( 3 - m \right) }
   {\left( 7 - m \right) }
\tilde{h}^{-\frac{2\,\left( 1 + m \right) }{m-1 }}+
  \frac{1 + m}{\left( 7 - m \right) } \tilde{h}^{-\frac{8}{3}}
\ee
(see Fig \ref{hofchi}).
In particular, for the point explosion case,
$m=3$,  this gives
\be
\tilde{h} = \chi^{-3/8}, \hskip .3 truein
\tilde{f} = \chi^{-1/2}, \hskip .3 truein
\tilde{g} = \chi^{-3/4}, \hskip .3 truein
\tilde{n} = \chi^{-3/8}
\label{31}
\ee
The local magnetization parameter $\beta= \tilde{h}^2/ \tilde{f} =\chi^{-1/4}$
is a slowly decreasing function of $\chi$.

Relations (\ref{31}) represent 
new self-similar solutions for strong  point  explosion in a 
strongly magnetized medium with $ 1 \ll \sigma \ll \Gamma^2 /4$.

As a test to these solutions we note (following the
  arguments of B\&M) that for  
point explosion  the energy associated with some interval of $d \chi$ should
remain constant. This leads to  the condition
\be
T^{0r} = T^{00} \beta_N
\ee
where $\beta_N = \left( { \partial r \over \partial t} \right) _{\chi = const}
$.
In the leading orders of $ 1 \ll \sigma \ll \Gamma^2$ this requires
\be
\chi = {  \tilde{f} \over  \tilde{g} \tilde{h}^2}
\ee
which is satisfied by the relations (\ref{31})

The total energy contained inside the relativistic strongly magnetized shock
is, for $m=3$,
\be
E\approx  4 \pi t^3 \int_0^R  \gamma^2 h^2 r^2 dr = {\pi \over 4}
 t^3 n_1 m_i c^3 y 
\ee
which is independent of time. This relation
 provides the  normalization for  $y=\Gamma^2$.

For the case of a  blast wave with  an energy supply we find:
\be
E = C(m) \, n_1 m_i c^3 t^3 y^{4/(1+m)} \sim t^{(3-m)/(m+1)}
\ee
with the constant of proportionality $C(m)$ being a complicated function
of $m$.
If the energy source is a power law function of time $L= L_0 t^q$, then
\be
m= { 2-q \over 2+q}
\ee
(c.f. B\&M Eq. (57)).

\section{Blast wave in an inhomogeneous medium}

In this section we assume the unshocked density and magnetic field
have power law dependences on the radius: $n_1 \sim r^{-k}$ and
$b_1 \sim r^{-l}$. As a classical example the non-relativistic constant velocity
wind gives $k=2, \, l=1$. 
Straightforward calculations give
\ba
&&
{{\cal{A}} \over g} { \partial \ln f \over  \partial \chi}= 
-2\,f\,\left( 4 + m\,\left( \chi g -4 \right)  + k\,\left(\chi g -2  \right)  - 4\, \chi g \right)  + 
\nn &&
\frac{h^2\,\left( -2\,\left( l + m-1 \right)  + 
        \left( -6 + 3\,k - 2\,l + 3\,m \right) \, \chi g \right) }
{ \chi g -1}
\nn &&
{{\cal{A}} \over g} { \partial \ln g \over  \partial \chi}= 
 f\,\left( -4 + 3\,k + 7\,m - 2\,\left( 2 + m \right) \, \chi g \right)+
  3\,h^2\,\left( l + m-1 \right)
\nn &&
{{\cal{A}}\over g} { \partial \ln h  \over  \partial \chi}= 
-\frac{f\,\left( -8 + 3\,k + 4\,l + 11\,m + \left( 8 + 3\,k - 16\,l - 11\,m \right) \, \chi g + 
        2\,\left( -4 + 2\,l + m \right) \, \chi^2 g^2 \right) }{2\,\left(  \chi g -1 \right) } + 
\nn &&
\frac{3\,h^2\,\left( l + m-1 \right) }{2}
\nn &&
{{\cal{A}}\over g} { \partial \ln n \over  \partial \chi}=
\frac{f\,
      \left( 12 - 7\,k - 11\,m + \left( -24 + 13\,k + 11\,m \right) \, \chi g - 
        2\,\left( -6 + 2\,k + m \right) \, \chi^2 g^2 \right) }{2\,\left(  \chi g -1 \right) }-
\nn &&
\frac{3\,h^2\,\left( l + m-1 + \left( 3 - 2\,k + l - m \right) \, \chi g \right) }
    {2\,\left(  \chi g -1 \right) }
\left.
\phantom{ {a\over b}}
\right)
\nn &&
{\cal{A}}  = \left( 1 + m \right) \,
\left(  2\,f\,\left( 1 - 4\, \chi\, g  + {\chi }^2 \, g^2 \right) -3\, \chi \,g\, h^2  \right)
\label{151}
\ea

In the limit of small $\sigma$ we reproduce equations (62-64)
of B\&M ($l$ naturally falls out) plus an equation for the magnetic field:
\be
{1\over \tilde{g}} {\partial \ln \tilde{h} \over \partial \chi}=
-\frac{\left( -16 + 6\,k+ 8 l+ 22\,m + \left( 8 +3\,k - 16 \, l - 11\,m \right) \,\chi \tilde{g} + 
      \left( -4 +2l + m \right) \,{\chi }^2 \, \tilde{g}^2 \right) }{2\,\left( 1 + m \right) \,
    \left( \chi \, \tilde{g} -2 \right) \,\left( 4 - 8\,\chi \, \tilde{g}  
+ {\chi }^2 \, \tilde{g}^2 \right) }
\ee
which for the point explosion case $m=3-k$ (magnetic energy density is not 
important) gives
\be
\tilde{g}= {1\over \chi},\hskip .3 truein
\tilde{h} = {\chi}^{\frac{k - 2\,\left( 4 - l \right) }{2\,\left( 4 - k \right) }} ,\hskip .3 truein
\tilde{f}= \chi^{(17 - 4 k)/(3k -12)} ,\hskip .3 truein
\tilde{n}=\chi^{(10-3k)/(4 (k-4))}
\ee
If the upstream medium has a constant magnetization parameter, then
$l=k/2$ and $ \tilde{h} = 1/ \chi$.

In the large $\sigma$ limit we find 
$
\tilde{g}=h^2 , \,
\tilde{f}= h^{4/3} , \,
\tilde{n}= \tilde{h} 
$
and  an equation for the magnetic field
\be
 {1\over \tilde{g}} {\partial \ln \tilde{h} \over \partial \chi}=
\frac{3\,\tilde{h}\,\left( l + m-1 \right) }
  {2\,\left( 1 + m \right) \,\left( \tilde{f} - 3\,\tilde{h}^2\,\tilde{g}\,\chi  \right) }
\ee
which may be resolved for $\chi(h):$
\be
\chi = {1 \over 7 - 2l -m}
\left((1 + m) \tilde{h}^{ - \frac{8}{3} } + 2\,\left( 3 -  2 l - m \right) 
\tilde{h}^{\frac{2\,\left( 11 - 8\,l - 5\,m \right) }{3\,\left( l + m-1 \right) }}  \right)
\ee
Again, similarly to the homogeneous case,
 the point explosion, $m=3-2l$ corresponds to 
$
\tilde{h}={1 / \chi^{3/8}}
$
The case of constant energy source in a current-free wind ($l=1,\, m=0$)
gives $\tilde{g}=\tilde{f}=\tilde{h}=\tilde{n}=1$

Similarly to the homogeneous case
 we can obtain the solutions inside the magnetosheath 
for the case of  nonzero $k$ and $l$. 
We find that the width of the boundary layer in coordinates $x$ is
\be
\Delta x= {\sigma }^{\frac{12 - 3\,k - m}{2\,\left( -4 + 3\,k - 4\,l + m \right) }}
\ee
The structure of the magnetic field in the layer is 
determined implicitly as
\be
x-1 = {C_2\,{\left(2- h^2   \right) }^
      {\frac{3\,\left( -2 + l \right) }{-4 + 3\,k - 4\,l + m}}}{h^
     {\frac{-12 + 3\,k + m}{4 - 3\,k + 4\,l - m}}}
\label{sa}
\ee
(compare with (\ref{dd})). Relation (\ref{sa})  determines the
behavior of the magnetic field on the contact discontinuity.
For example, for a constant external magnetization parameter $l=k/2$ 
magnetic field completely dominates the boundary layer dynamics
for $k<4$ and $k+m<4$. For  a shock wave propagating in  a non-relativistic
constant velocity and constant magnetization wind ($k=2, l=1$) 
magnetic field goes to a constant while density and pressure go to
zero for $m < 2$.

\section{Discussion}

We have  analyzed the
propagation of a spherically symmetric relativistic
shock wave into an 
inhomogeneous medium permeated by  toroidal magnetic field.
The main limitation of the current work is the 
neglect of the possible lateral expansion.
The natural and much more complicated
  extension of this approach would include a 
blast wave in a constant magnetic field. 

The self-similar solutions presented here are  generally 
applicable to the expansion
of a   strongly relativisticly blast wave in a preexisting 
spherical 
cavity blown
by the magnetized  wind
and in a equatorial region of a constant external magnetic field.
The  self-similar 
structure of a thin magnetosheath, on the other hand, has a much broader
validity  since   any radial (orthogonal to the surface of the
contact discontinuity) magnetic field should approach 0 on the surface
of the CD.

The results of this work bear relevance to the physics of 
Gamma Ray Bursts. 
In particular, the most popular model of GRBs relies on  relativisticly
strong shock waves to produce both Gamma-ray and afterglow emission
\cite{Piran00}. Previously, the dynamics of the magnetic field has not
been included in the model.
In addition, the very basic assumptions of the model - presence of 
near-equipartition magnetic fields and acceleration of particles by
relativistic shocks - have been criticized \cite{GallantAchterberg99}.
Our results point to the new interesting possibility that the  relativistic flow
 may produce radiation effectively in the  magnetosheath layer adjacent to the
contact discontinuity between two flows
where magnetic field reaches near-equipartition values. We leave the 
investigation of this possibility to a future work.


\acknowledgments
I would like to thank Roger Blandford, Vladimir Usov, Michail Medvedev
and Malory Roberts for useful comments and suggestions.

\appendix

\section{Dynamic equations for independent variable  $x= g\chi$ }

Changing to the new variable $x = g\chi$ in (\ref{15})
and using
\be
dx= \frac{\left( -3\,h^2\,\left( -2 + l \right) \,x + 
      f\,\left( -2 + m\,\left( x-2 \right)  - 3\,\left( k-4 \right) \,x + 2\,x^2 \right) 
      \right) \,g(x)}{\left( 1 + m \right) \,
    \left( 3\,h^2\,x - 2\,f\,\left( 1 - 4\,x + x^2 \right)  \right) } \, d\chi
\ee
we find
\ba
 (x-1)\, {\cal{A'} }\, { \partial \ln f \over  \partial x } &&=
2\,f\,\left( 4 + m\,\left( -4 + x \right)  + k\,\left( x-2 \right)  - 4\,x \right) \,
  \left( x-1 \right)
\nn &&
+ h^2\,\left( 2\,\left( l + m-1 \right)  + \left( 6 - 3\,k + 2\,l - 3\,m \right) \,x \right)
\nn 
{\cal{A'} }\, { \partial \ln g\over  \partial x} &&= 
f\,\left( 4 - 3\,k - 7\,m + 2\,\left( 2 + m \right) \,x \right)
-3\,h^2\,\left( l + m-1 \right)
\nn
(x-1)\, {\cal{A'} }\, { \partial \ln h \over  \partial x}&&= 
{1\over 2} f\,\left( -8 + 3\,k + 4\,l + 11\,m + \left( 8 + 3\,k - 16\,l - 11\,m \right) \,x + 
    2\,\left( -4 + 2\,l + m \right) \,x^2 \right)-
\nn &&
{3\over 2} h^2\,\left( l + m-1 \right) \,\left( x-1 \right)
\nn 
 (x-1)\, {\cal{A'} } \, { \partial \ln n \over  \partial x} &&=
{1\over 2} \left(-12 + 7\,k + 11\,m + \left( 24 - 13\,k - 11\,m \right) \,x + 
  2\,\left( -6 + 2\,k + m \right) \,x^2 \right) f +
 \nn &&
{3 \over 2}  \left( l + m-1 + \left( 3 - 2\,k + l - m \right)\,x \right) h^2
\nn 
{\cal{A'} } &&=-3\,h^2\,\left( -2 + l \right) \,x + f\,
   \left( -2 + m\,\left( x-2 \right)  - 3\,\left( k-4 \right) \,x + 2\,x^2 \right)
\label{5}
\ea

In coordinate $x$ the shock is located at $x_0= g_0$ and the contact discontinuity is located at $x=1$.

\section{Lower dimension outflows}

Without giving detail calculations, we comment here on the
generalization of the above results for the lower dimensional systems,
when the outflow is cylindrically or linearly symmetric.
We  give here only the scalings for the
 point explosion cases in a homogeneous medium, 
generalizations for inhomogeneous media with energy supply are straightforward.
For 1-D case (strong shock propagating along a tube) the solutions
look very similar to the 3-D case. In the small $\sigma$ limit
we find $ \tilde{g} =  \tilde{h}= \tilde{n}=1/\chi, \, \tilde{f} = \chi^{-7/6}$,
while for the large $\sigma$ limit 
we reproduce relations (\ref{31}). The 
structure of the equations in the 2-D case is, in fact, {\it qualitatively}
different from the 1-D and 3-D cases. In
the small $\sigma$  limit:
 $ \tilde{g} =1/\chi, \,   \tilde{h}=1/\chi^{5/6}, \, \tilde{n}=1/\chi^{7/6}, 
\, \tilde{f} = \chi^{-4/3}$,
 while the large $\sigma$ turns out to be unusual:
$\tilde{g} =  \chi^{-5/6}, \, \tilde{n}= \tilde{h}= \chi^{-5/12}, \,
  \tilde{f} =  \chi^{-2/3}$.

\newpage

\end{document}